\documentclass[conference]{IEEEtran}
\IEEEoverridecommandlockouts
\let\svtilde~

\newcommand\newtildeON[1][A]{\def~{\csname newtilde#1\endcsname}}
\newcommand\newtildeOFF{\let~\svtilde}
\usepackage{cite}
\makeatletter 
\newcommand{\linebreakand}{%
  \end{@IEEEauthorhalign}
  \hfill\mbox{}\par
  \mbox{}\hfill\begin{@IEEEauthorhalign}
}
\makeatother 

\usepackage{amsmath,amssymb,amsfonts}
\usepackage{algorithmic}
\usepackage{graphicx}
\usepackage[bottom]{footmisc}
\usepackage{textcomp}
\usepackage{listings}
\usepackage[table]{xcolor}
\usepackage{mathtools}
\usepackage{diagbox}
\usepackage{enumitem} 
\usepackage{subfigure}
\usepackage{graphicx}
\definecolor{codegreen}{rgb}{0,0.6,0}
\definecolor{codegray}{rgb}{0.5,0.5,0.5}
\definecolor{codepurple}{rgb}{0.58,0,0.82}
\definecolor{backcolour}{rgb}{0.95,0.95,0.92}

\definecolor{dkgreen}{rgb}{0,0.6,0}
\definecolor{gray}{rgb}{0.5,0.5,0.5}
\definecolor{mauve}{rgb}{0.58,0,0.82}

\def\BibTeX{{\rm B\kern-.05em{\sc i\kern-.025em b}\kern-.08em
    T\kern-.1667em\lower.7ex\hbox{E}\kern-.125emX}}
\begin{document}

\title{Adaptive Background Music for a Fighting Game: A Multi-Instrument Volume Modulation Approach
}

\author{\IEEEauthorblockN{Ibrahim Khan}
\IEEEauthorblockA{\textit{Graduate School of Information Science and Engineering} \\
\textit{Ritsumeikan University}\\
Kusatsu, Shiga, Japan \\
gr0556vx@ed.ritsumei.ac.jp}

\and
\IEEEauthorblockN{Thai Van Nguyen}
\IEEEauthorblockA{\textit{Graduate School of Information Science and Engineering} \\
\textit{Ritsumeikan University}\\
Kusatsu, Shiga, Japan \\
gr0557fv@ed.ritsumei.ac.jp}

\linebreakand 
\IEEEauthorblockN{Chollakorn Nimpattanavong}
\IEEEauthorblockA{\textit{Graduate School of Information Science and Engineering} \\
\textit{Ritsumeikan University}\\
Kusatsu, Shiga, Japan \\
gr0608sp@ed.ritsumei.ac.jp}

\and

\IEEEauthorblockN{Ruck Thawonmas}
\IEEEauthorblockA{\textit{College of Information Science and Engineering} \\
\textit{Ritsumeikan University}\\
Kusatsu, Shiga, Japan \\
ruck@is.ritsumei.ac.jp}
}

\maketitle

\begin{abstract}
This paper presents our work to enhance the background music (BGM) in DareFightingICE by adding an adaptive BGM. The adaptive BGM consists of five different instruments playing a classical music piece called “Air on G-String.” The BGM adapts by changing the volume of the instruments. Each instrument is connected to a different element of the game. We then run experiments to evaluate the adaptive BGM by using a deep reinforcement learning AI that only uses audio as input (Blind DL AI). The results show that the performance of the Blind DL AI improves while playing with the adaptive BGM as compared to playing without the adaptive BGM.
\end{abstract}


\section{Introduction}
As the popularity of video games keeps growing, video game developers are trying to improve their games to leave a positive impression on the players by enhancing their experience \cite{b1}. These improvements are in many different forms like gameplay-related changes or visually pleasing games or games with very enjoyable music or sound effects. Regardless of the type of improvement, the ultimate goal is to enhance the overall experience of the players.\par

Our work is based on DareFightingICE Competition, which has two tracks: Sound Design Track and AI Track \cite{b2}. It ran at the 2022 IEEE Conference on Games (CoG 2022). The platform for the competition is DareFightingICE -- an enhanced version of FightingICE \cite{b3} -- with an enhanced sound design and a new AI interface that can give audio data to AIs. \par

This research focuses on video game music and how to use it to enhance players’ experience. An immersive gaming experience is greatly aided by BGM \cite{b5,b6}. In fighting games, BGM should change in response to the dynamic action, creating suspense and excitement as the players engage in combat. Fighting game BGM is typically implemented using pre-composed songs that loop, which fall short of accurately capturing the intensity of combat.\par

As a result, the target genre in this research is fighting games. In fighting games, players go against another player or a computer player in a one-versus-one fight, using different attacks and abilities to overcome the opponent. These fighting games are two-and-a-half-dimensional, which means that the players can only move in two dimensions (left or right, while in some fighting games the players can perform an attack or dodge that takes them out of 2D). In addition, because fighting games are fairly simple compared to other genres of video games, they are open to a larger audience.  This research uses DareFightingICE.\par

The goal of this research is to create adaptive BGM by modifying the BGM of DareFightingICE, and evaluating the performance of the adaptive BGM by using a deep learning AI that only uses audio as input (Blind DL AI) \cite{b4}. The contributions of our work are as follows:
\begin{enumerate} \item One we are the first group to focus on adaptive music in fighting games. \item we are also the first group to use multiple instruments in the adaptation of a BGM. \item our research focuses on giving players information about the state of the game through the BGM.\par
\end{enumerate}

\section{Related Work}
Since our research touches on two topics which are audio in video games and adaptive audio in video games, this section is divided into two parts.
\subsection{Sound in Video games}

The significance of music in video games has been extensively studied. Video game music has been proven to have an impact on everyday living, attitude, and other aspects \cite{b7}. A person's attachment to the music featured in a game can result in these effects. It has been established that voice-over in video games and audio dialogue acted by the player or by non-playable characters is more entertaining for the players \cite{b8}. The information given to the game's players is also made easier to recall thanks to these voiceovers. A significant part of the game is also performed by the background soundtrack. It was found that players did better when there was background music playing than when there wasn't any \cite{b9}. \par

Similarly, it has been noted that for video games to be engaging for players, the music performed within them must match the atmosphere or tone of the game \cite{b10}. A poor illustration of this would be to play soothing music as the game reaches its conclusion. Moving from background music to more focused sound effects and auditory signals, it has been observed that directional or 3D sound effects give players more information about their surroundings, including where other players are located \cite{b11,b12}. Steps and other types of action are represented in these 3D sound elements.\par

This section's focus thus far has been on video games' audio. There has been very little study on sound designs in video games, which is a change from sound in video games. A sound design refers to a collection of sound effects together with its source code for its timing-control algorithm. Even fewer have produced findings that are helpful in developing an effective sound design. Despite the lack of study, sound designs for game categories like first-person shooters and real-time strategy have been suggested \cite{b13}. For their respective game categories, these two sound designs have distinctive sound effects.\par

\subsection{Adaptive Audio in Video Games}

There has been some research on adaptive music in video games in the past. A study made an adaptive music system keeping players’ emotions in mind and it was shown to be better than the original music in the game \cite{b14}. Another study on adaptive music found that players value adaptive music over linear music \cite{b15}. Lastly, another research on adaptive music changed the tempo of the music by taking into account players’ actions and the state of the game and found that it enhanced players’ experience by making the game more immersive or enjoyable \cite{b16}.\par

Previous research has shown the importance of both music and adaptive music in video games. However, there has been little to no research when it comes to adaptive music in fighting games.\par

\subsection{DareFightingICE Platforms}
The adaptive BGM is created in the aforementioned fighting game platform DareFightingICE. Since there has been a recent update to the DareFightingICE platform from version 5.2 (used in the 2022 DarefightingICE Competition) to version 6.0, we include both versions in our research. The reason is to show that our adaptive BGM works well on both of them.

\subsubsection{DareFightingICE 5.2}
DareFightingICE version 5.2 was the official version of the DareFightingICE Competition which was first held at CoG 2022. This version was an upgrade from the FightingICE platform and added an enhanced sound design keeping players without vision in mind. The platform also provided audio data to AIs, leading to the new Blind AI Track of the competition.\par

Version 5.2 added a new function getAudioData to the original interface that gives audio data at each frame of the game to make it easier for AIs to access audio data. At each frame, audio data are sampled with a length of 16.67 ms; notice that DareFightingICE has a frame per second of 60, as in FightingICE. The stereo sound in the 2D format used in DareFightingICE helps accurately depict the location of both players. Consequently, a two-channel sound format is offered to enable AIs to perceive audio that might aid them in locating the positions of both players and those of projectiles. \par

\subsubsection{DareFightingICE 6.0}

This version of DareFightingICE has revamped the communication interface between AIs and the game system by using the open-source remote procedure call gRPC instead of Py4J. This change has resulted in up to 65\% reduced latency, improved stability, and the elimination of missed frames compared to the Py4J interface \cite{b17}. This version will be used in the 2023 DareFightingICE Competition with another editable source code file called “Play.java” included for the Sound Design Track. This file is made available to give the users ability to modify the BGM at run time.

\section{Rule-Based Adaptive BGM}
\begin{figure}[t!]
\centerline{\includegraphics[width=0.5\textwidth]{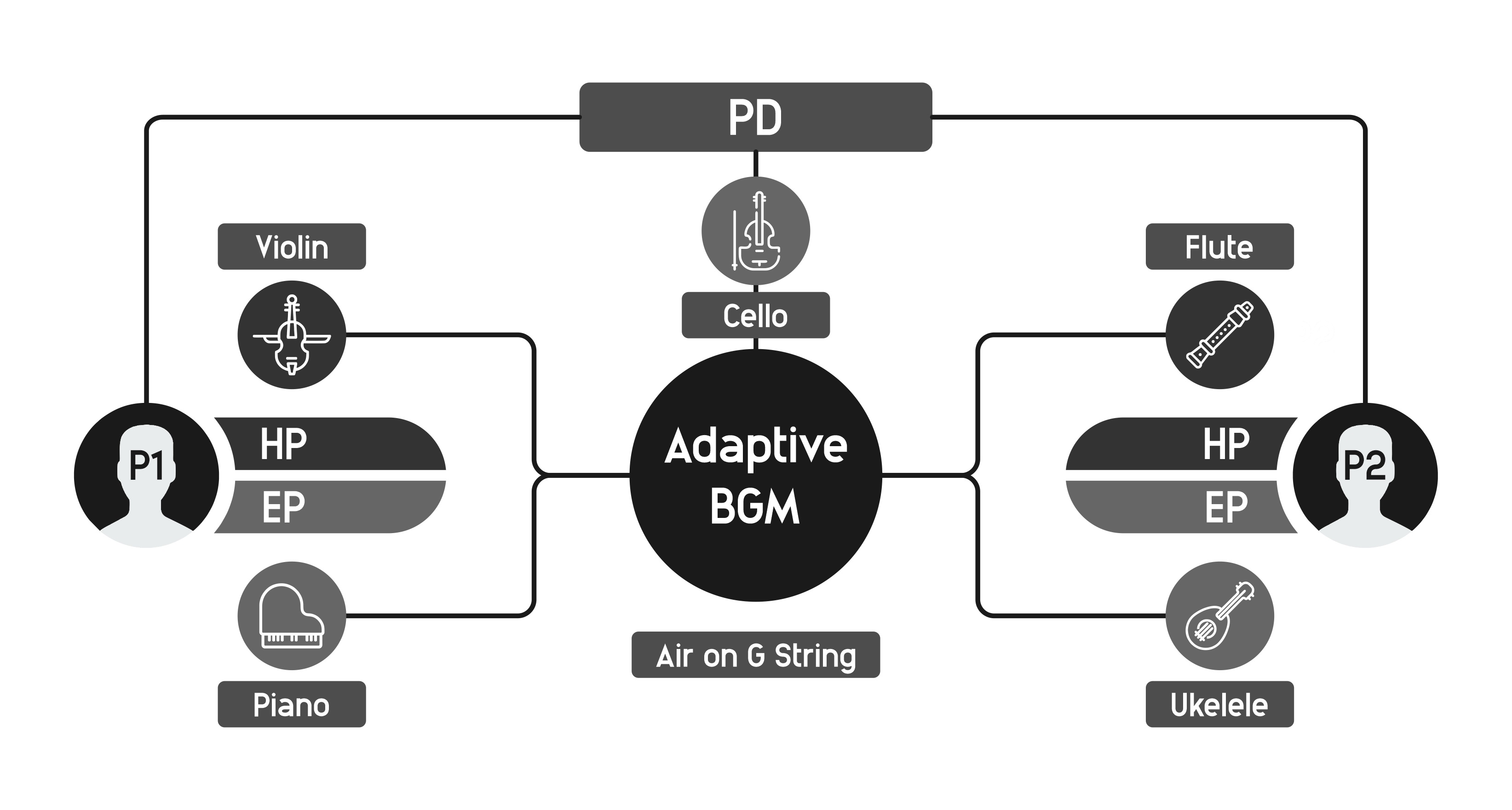}}
\caption{Rule-Based Adaptive BGM.}
\label{figADM}
\end{figure}

In this research, we propose an adaptive BGM that adapts to player actions and the players' in-game position. The proposed adaptive BGM consists of five different instruments playing a classical music piece called “Air on G-String.” This music was selected for this research because we found the contrast of fast-paced actions with calming music to be a good combination, a technique used in some famous and popular movies such as X-Men: Days of the Future Past (QuickSilver-stopping-bullets scene). The five different instruments are the piano, the cello, the flute, the violin, and the ukulele. The BGM adjusts by altering the loudness of the instruments. This adjustment happens when the game elements, these instruments are connected to, change. The elements in use are both players’ health points (HP), energy points (EP), and the distance between the two players (PD). The HP is the number of hits a player can take before losing, and the EP is the energy in point that the players need to perform different attacks. The design of the proposed adaptive BGM is illustrated in Fig.~\ref{figADM}. \par

In Fig.~\ref{figADM}, there are five instruments playing together to compose the BGM, and every instrument is connected to a different game element. The violin and the piano are connected to player one’s HP and EP, respectively; and the flute and the ukulele are connected to player two’s HP and EP, respectively. Lastly, the cello is connected to PD.\par

\begin{figure}[b!]
\begin{lstlisting}
For Violin:
    if (P1.HP decreases)
        Violin.volume.decreaselevel
For Flute:
    if (P2.HP decreases)
        Flute.volume.decreaselevel
For Cello:
    if(Pos.diff decreases)
        Cello.volume.increaselevel
    Else
        Cello.volume.decreaselevel
For Piano:
    if(P1.EP decreases)
        Piano.volume.decreaselevel
    Else
        Piano.volume.increaselevel
For Ukulele:
    if(P2.EP decreases)
        Ukulele.volume.decreaselevel
    Else
        Ukulele.volume.increaselevel
\end{lstlisting}
\caption{Adaptive BGM Rules}
\label{figAIInterface}
\end{figure}
In Fig.~\ref{figAIInterface}, the instruments connected to HP of both players are at maximum volume when HP is maximum, and the volume is turned down as the HP gets lower. The ones connected to players’ EP work in the same way, i.e., when EP is low the volume is low and when EP is high the volume is high. For the cello, the volume is at maximum when PD is close to zero meaning they are very close to each other, and it becomes lower the further away the players move from each other. The proposed adaptive BGM is designed this way to give useful information to both human and AI players.\par

The maximum HP for one player per round is 400, EP is 300, and the PD is 800 pixels horizontally. The HP and EP of both players and their PD are divided into different levels to decrease or increase the volume of the instruments gradually. The levels of not only instruments but also players’ HP, EP, and PD are empirically selected due to the lack of existing research when it comes to adaptive BGM using volume modulation.\par

The levels for HP are 400, 300, 250, 200, 150, 100, and 50, which are connected to the instrument’s volume levels of 75\%, 60\%, 55\%, 40\%, 35\%, 25\%, and 10\% respectively. For  EP of both players, the instruments’ volume levels are the same as HP, but EP levels are 300, 250, 200, 150, 100, and 50. For PD, the levels are 800, 600, 500, 400, 300, 60, and 0,  which are connected to the instrument’s volume levels of 10\%, 20\%, 30\%, 40\%, 50\%, 60\%, and 75\%. The reason for the abrupt change from 300 to 60 is for the players without vision to know without a doubt that the opponent player is near them.

\begin{figure}[b!]
\begin{lstlisting}
  public void playBGM() {
    play2(this.Violin, getSoundBuffers().get("bach_V.wav"), 350, 0, true);
    play2(this.Piano, getSoundBuffers().get("bach_P.wav"), 350, 0, true);
    play2(this.Flute, getSoundBuffers().get("bach_F.wav"), 350, 0, true);
    play2(this.Ukulele, getSoundBuffers().get("bach_U.wav") , 350, 0, true);
    play2(this.Cello, getSoundBuffers().get("bach_C.wav"), 350, 0, true);

    }
\end{lstlisting}
\caption{Adaptive BGM Code Example}
\label{figInterface}
\end{figure}

Lastly, for the 2023 DareFightingICE Competition participants, if they want to modify or change the adaptive BGM\footnote{code of the adaptive BGM: http://tinyurl.com/Adapt-BGM}. They just have to change the five instruments playing the BGM and if they want to use fewer than five instruments, they can decide which rules to ignore and which to select. Figure~\ref{figInterface} shows a code snippet that they can modify. In Fig.~\ref{figInterface}, the play2() function takes five parameters as follows:
\begin{enumerate}
\item The first parameter is the audio source of the instrument.
\item The second parameter is the audio buffer that stores the audio file that will play on the audio source.
\item The third parameter is the horizontal location in the game, where the audio source will play.
\item The fourth parameter is the vertical location in the game. \item The last parameter is to check if the audio source should be looped or not.
\end{enumerate}

\section{Evaluation}
For the evaluation of the adaptive BGM, we did an objective evaluation, the details for which are following.

For the objective evaluation, we use the aforementioned Blind DL AI. Our method for this evaluation is to train the Blind DL AI with and without the adaptive BGM and then compare the performance of the AIs. We hypothesize that if the performance of the Blind DL AI is better with the adaptive BGM then the adaptive BGM is giving useful information, and since the AI we are using can only take audio data as input we believe that players without vision will also be able to use the same information for their advantage.\par

To test our adaptive BGM we used the submissions to the 2022 DareFightingICE Competition Sound Design Track and train the Blind DL AI as mentioned above. We trained the Blind DL AI for 900 rounds on DareFightingICE 5.2 (1 game has 3 rounds) against MCTSAI65, a weaken version of a sample Monte-Carlo \cite{b18}. For DareFightingICE 6.0, we implement a new weak version of MCTS AI, which fixed the problem of MCTSAI65's performance difference on different environments, and call it MCTSAI23i. This version of MCTS AI has similar performance as the previous MCTSAI65 against our Blind DL AI as shown in Table ~\ref{tblFightResults1}. We trained the Blind DL AI against the MCTSAI23i for 900 rounds for each sound design.\par

\begin{table}[t!]
\caption{Performance of our blind DL AI against MCTSAI65 and MCTSAI23i.}
\label{tblFightResults1}
\begin{center}
\begin{tabular}{|c|c|c|c|}
\hline
 Opponent & Encoder & $win_{ratio}$ & $avg HP_{diff}$\\
\hline
MCTSAI65 &1D-CNN & 0.33 & -28.83\\
\hline
MCTSAI65 &FFT & 0.37 & -40.5\\
\hline
MCTSAI65 &Mel-spectrogram &  0.63 &  37.07\\
\hline
MCTSAI23i &1D-CNN &  0.53 &  12.92\\
\hline
MCTSAI23i &FFT & 0.37 & -30.08\\
\hline
MCTSAI23i &Mel-spectrogram &  0.54 &  28.18\\
\hline
\end{tabular}
\end{center}
\end{table}

The difference between MCTSAI65 and MCTSAI23i is that for MCTSAI65 the MCTS execution time was changed to 6.5 ms and for MCTSAI23i we limit the number of iterations MCTS can do for each frame. We introduce this change because we found that MCTSAI65 performed differently in different environments.\par

For the subjective evaluation or evaluation of our adaptive BGM by human players, since the performance of Blind DL AI was similar to the performance of human players in previous work \cite{b2}, we decided to only go with the objective evaluation. We believe that both players with and without vision will also be able to use the same information to their advantage.

We used the same environments as in the 2022 DareFightingICE Competition. More specifically, six computers were used that have the same specification, i.e., CPU: Intel(R) Xeon(R) W-2135 CPU@ 3.70GHz 3.70 GHz, RAM: 16 GB, GPU: NVIDIA Quadro P1000 4GB VRAM, and OS: Windows 10.\par

\subsection{Results}

We conducted experiments on three different sound designs; The default, runner-up, and winner sound design of the 2022 DareFightingICE Competition on both versions of DareFightingICE.\par

The experiment structure is that the Blind DL AI fights against MCTSAI65 for version 5.2 and MCTSAI23i for version 6.0 for each sound design, first with linear BGM and then with the BGM being replaced with the adaptive one. The AI is trained from scratch for 900 rounds and since the Blind DL AI had three different audio encoders \cite{b3} as an option, the experiment was run on each encoder per one sound design; one-dimensional Convolutional Neural Network (1D-CNN), Fast Fourier Transform (FFT), and Mel-Spectrogram (Mel-Spec).\par

We then evaluate the performance of each trained Blind DL AI by making it fight against each of the aforementioned opponent AIs for 90 rounds. The ratio of the number of wins\footnote{In the game, the round winner is either the one with a non-zero HP while its opponent's HP has reached zero or the one with the higher HP when the round-length limit of 60 s has reached.} over 90 rounds, Eqn. \eqref{equWinRate}, and the average HP difference at the end of a round between the trained AI and its opponent, Eqn. \eqref{HPDifference}, are then calculated. The equations and details above are taken from previous work \cite{b4}. \par
\begin{equation}
    win_{ratio} = \frac {\textit{winning rounds}}{\textit{total rounds}}\label{equWinRate}
\end{equation}
\begin{equation}
    avg HP_{diff} = \frac {\textit{sum of }HP_r^{self} - HP_r^{opp}\textit{ for all r}}{\textit{total rounds}} \label{HPDifference}
\end{equation}

\begin{table}[t!]
\caption{Performance of our blind DL AI with default sound design in DareFighingICE 5.2.}
\label{tblFightResults2}
\begin{center}
\begin{tabular}{|c|c|c|c|}
\hline
Sound design &Encoder & $win_{ratio}$ & $avg HP_{diff}$\\
\hline
Default &1D-CNN & 0.33 & -28.83\\
\hline
Default &FFT & 0.37 & -40.5\\
\hline
Default &Mel-spectrogram &  0.63 &  37.07\\
\hline
Default Adaptive &1D-CNN & {\bf 0.79} & {\bf 81.98}\\
\hline
Default Adaptive &FFT &{\bf 0.92} & {\bf118.93}\\
\hline
Default Adaptive &Mel-spectrogram &{\bf  0.87} &{\bf  105.61}\\
\hline
\end{tabular}
\end{center}
\end{table}
\begin{table}[t!]
\caption{Performance of our blind DL AI with runner-up sound design in DareFighingICE 5.2.}
\label{tblFightResults4}
\begin{center}
\begin{tabular}{|c|c|c|c|}
\hline
Sound design &Encoder & $win_{ratio}$ & $avg HP_{diff}$\\
\hline
Runner-up &1D-CNN & 0.6& 42.91\\
\hline
Runner-up &FFT & 0.24 & -55.13\\
\hline
Runner-up &Mel-spectrogram &  0.0 &  -282.86\\
\hline
Runner-up Adaptive &1D-CNN & {\bf 0.92} & {\bf 102.78}\\
\hline
Runner-up Adaptive &FFT &{\bf 0.81} & {\bf82.04}\\
\hline
Runner-up Adaptive &Mel-spectrogram &{\bf  0.91} &{\bf  121.86}\\
\hline
\end{tabular}
\end{center}
\end{table}
\begin{table}[t!]
\caption{Performance of our blind DL AI with Winner sound design in DareFighingICE 5.2.}
\label{tblFightResults3}
\begin{center}
\begin{tabular}{|c|c|c|c|}
\hline
Sound design &Encoder & $win_{ratio}$ & $avg HP_{diff}$\\
\hline
Winner &1D-CNN & 0.47 & -5.73\\
\hline
Winner &FFT & 0.80 & 75.54\\
\hline
Winner &Mel-spectrogram &  0.0 &  -169.41\\
\hline
Winner Adaptive &1D-CNN & {\bf 1} & {\bf 142.61}\\
\hline
Winner Adaptive &FFT &{\bf 0.87} & {\bf112.31}\\
\hline
Winner Adaptive &Mel-spectrogram &{\bf  0.71} &{\bf  68.88}\\
\hline
\end{tabular}
\end{center}
\end{table}

\begin{table}[t!]
\caption{Performance of our blind DL AI with default sound design in DareFighingICE 6.0.}
\label{tblFightResults5}
\begin{center}
\begin{tabular}{|c|c|c|c|}
\hline
Sound design &Encoder & $win_{ratio}$ & $avg HP_{diff}$\\
\hline
Default &1D-CNN & 0.53 & 19.92\\
\hline
Default &FFT & 0.37 & -30.08\\
\hline
Default &Mel-spectrogram &  0.54 &  28.18\\
\hline
Default Adaptive &1D-CNN & {\bf 0.59} & {\bf 31.54}\\
\hline
Default Adaptive &FFT &{\bf 0.67} & {\bf29.59}\\
\hline
Default Adaptive &Mel-spectrogram &{\bf  0.64} &{\bf  41.22}\\
\hline
\end{tabular}
\end{center}
\end{table}
\begin{table}[t!]
\caption{Performance of our blind DL AI with runner-up sound design in DareFighingICE 6.0.}
\label{tblFightResults7}
\begin{center}
\begin{tabular}{|c|c|c|c|}
\hline
Sound design &Encoder & $win_{ratio}$ & $avg HP_{diff}$\\
\hline
Runner-up &1D-CNN & 0.56 & 31.68\\
\hline
Runner-up &FFT & 0.64 & 47.73\\
\hline
Runner-up &Mel-spectrogram &  0.60 &  35.02\\
\hline
Runner-up Adaptive &1D-CNN & {\bf 0.68} & {\bf 66.12}\\
\hline
Runner-up Adaptive &FFT &{\bf 0.77} & {\bf98.23}\\
\hline
Runner-up Adaptive &Mel-spectrogram &{\bf  0.61} &{\bf  56.19}\\
\hline
\end{tabular}
\end{center}
\end{table}
\begin{table}[t!]
\caption{Performance of our blind DL AI with winner sound design in DareFighingICE 6.0.}
\label{tblFightResults6}
\begin{center}
\begin{tabular}{|c|c|c|c|}
\hline
Sound design &Encoder & $win_{ratio}$ & $avg HP_{diff}$\\
\hline
Winner &1D-CNN & 0.54 & 12.66\\
\hline
Winner &FFT & 0.52 & 27.03\\
\hline
Winner &Mel-spectrogram &  0.56 &  34.23\\
\hline
Winner Adaptive &1D-CNN & {\bf 0.73} & {\bf 71.91}\\
\hline
Winner Adaptive &FFT &{\bf 0.66} & {\bf32.03}\\
\hline
Winner Adaptive &Mel-spectrogram &{\bf  0.65} &{\bf  45.62}\\
\hline
\end{tabular}
\end{center}
\end{table}

As shown from the results in Tables \ref{tblFightResults2},\ref{tblFightResults3},\ref{tblFightResults4},\ref{tblFightResults5},\ref{tblFightResults6}, and \ref{tblFightResults7}, the Blind DL AI performs better with the adaptive BGM in all sound designs across both versions of DareFighingICE. For each encoder, Blind DL AI also performs better with our adaptive BGM. This proves our hypothesis that the adaptive BGM is giving useful information. \par

For version 5.2 the best encoder is 1D-CNN for the winner sound design. It is undefeated in testing. This encoder is also the most stable and consistent among all other encoders for all the sound designs for version 5.2. \par

Version 6.0's best encoder is FFT for the runner-up sound design as it has the highest performance among all encoders for all sound designs. We call this encoder best for version 6.0 because it is also consistent in testing. \par

 \subsection{Behavior of Blind DL AIs}
 This section describes the behavior\footnote{Link to the videos of Blind DL AIs: https://tinyurl.com/abm4fg.} of different Blind DL AIs for different sound designs in both versions. Only the behavior of Blind DL AIs trained on the adaptive BGM is described. For each sound design, the encoder with the best result is chosen. The summary of different behaviors is shown in Table \ref{tblFightResults8}. 

\begin{figure*}
    \centering
    \includegraphics[width=.16\linewidth]{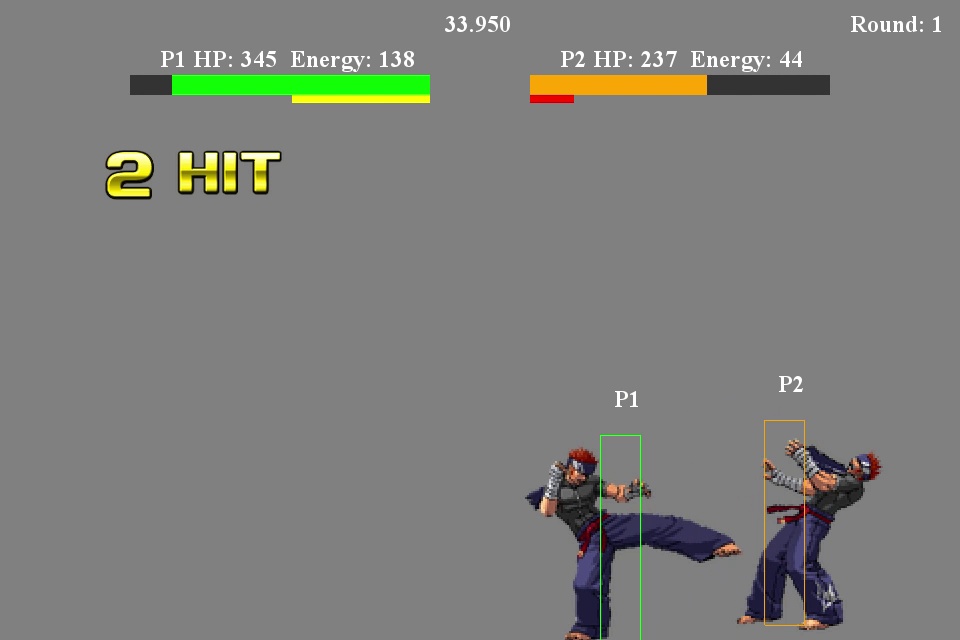}\hfill
    \includegraphics[width=.16\linewidth]{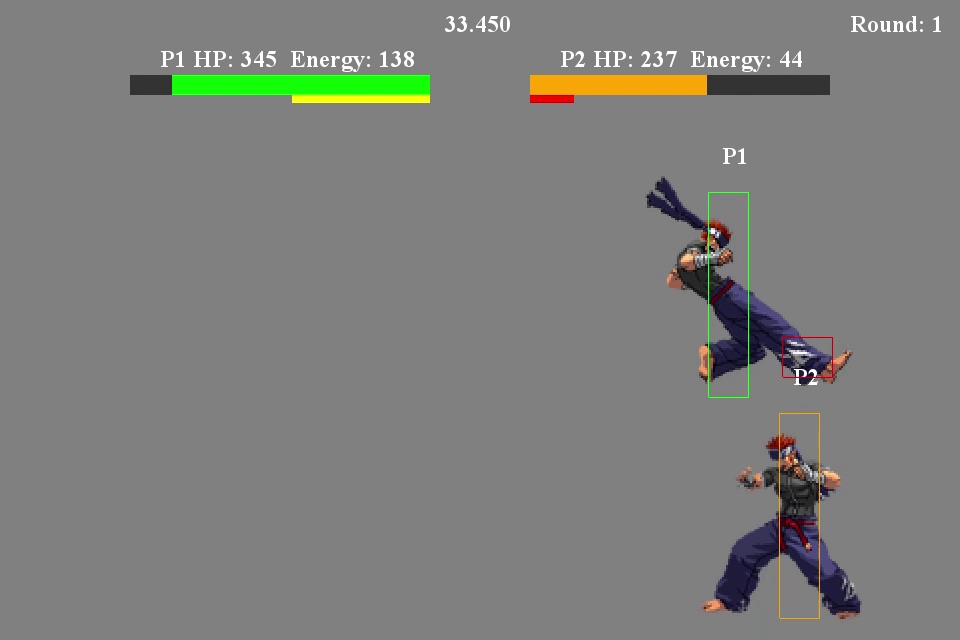}\hfill
    \includegraphics[width=.16\linewidth]{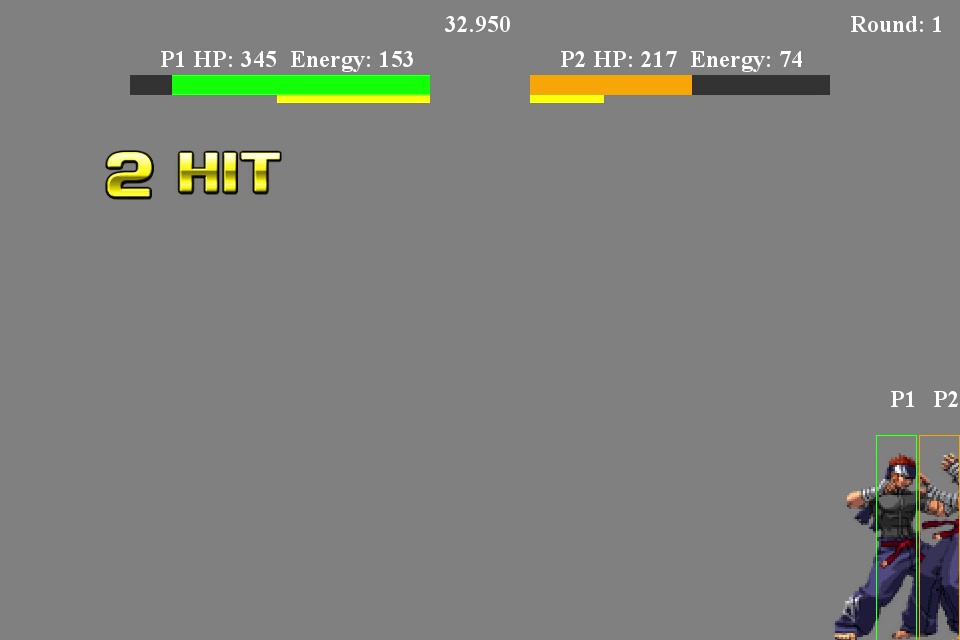}\hfill
    \includegraphics[width=.16\linewidth]{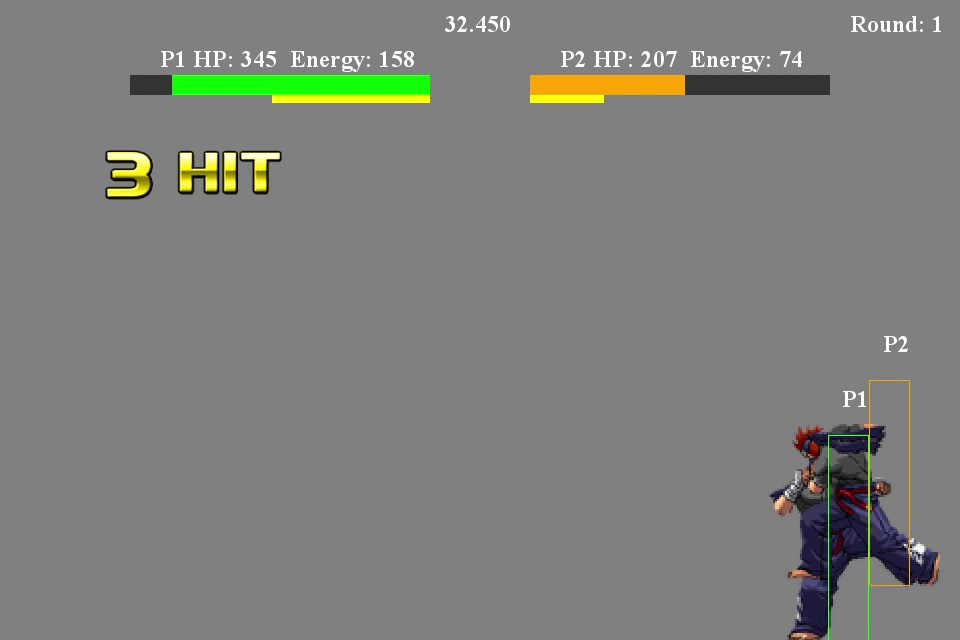}\hfill
    \includegraphics[width=.16\linewidth]{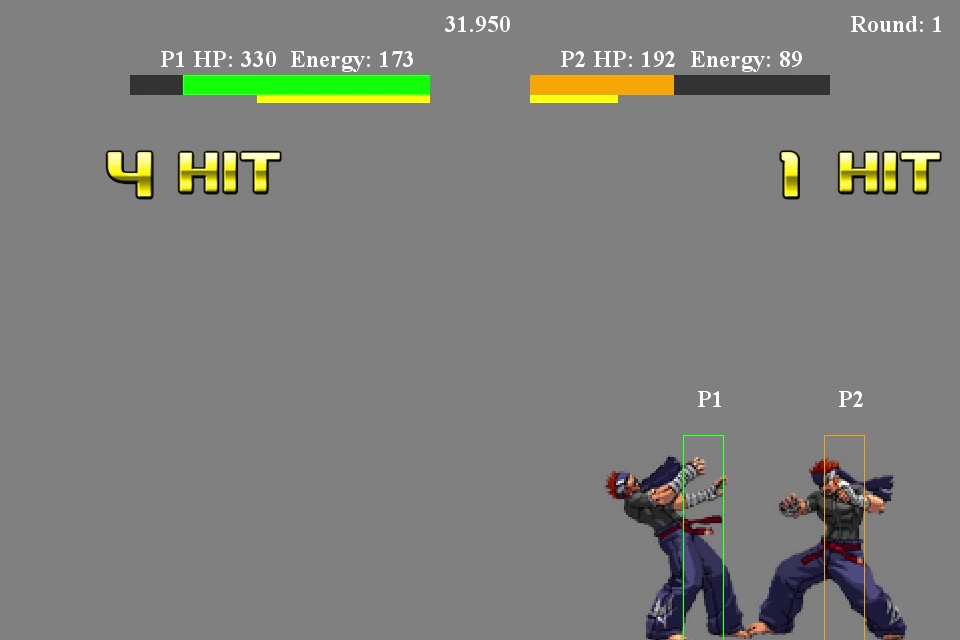}\hfill
    \includegraphics[width=.16\linewidth]{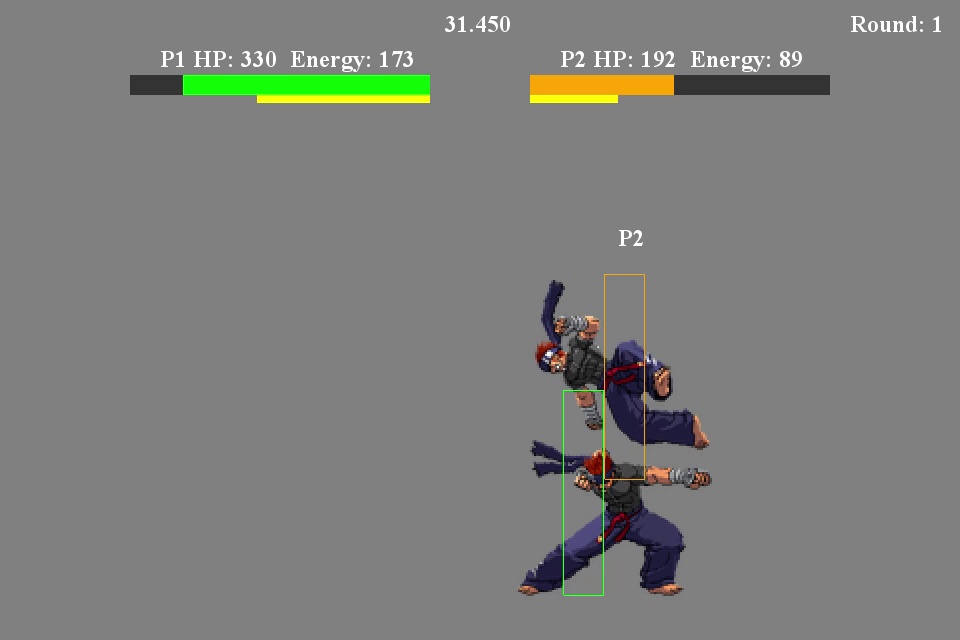}\hfill
    \includegraphics[width=1\linewidth]{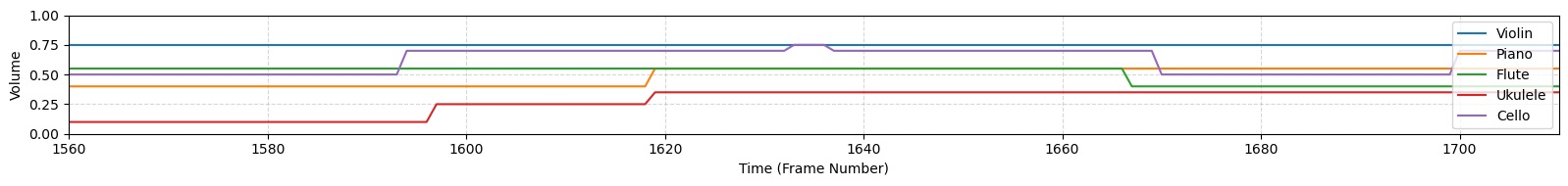}\hfill
    \caption{A series of fight scenes at different time steps and the level of different instrument volume at these time steps.}
    \label{fig:figFightingSample}
\end{figure*}

 Figure~\ref{fig:figFightingSample} shows a sample of different time steps from a fight between the best Blind DL AI from version 6.0 against MCTSAI23i, and it also shows the change in volume of different instruments at these time steps.
 
\begin{table}[t!]
\caption{Difference between the different Blind DL AIs.}
\label{tblFightResults8}
\begin{center}
\resizebox{\linewidth}{!}{%
\begin{tabular}{|c|c|c|c|c|c|}
\hline
Version &Sound Design & Playing Style & Favorite Move(s) & Avoid Projectiles & Lighting Uppercut \\
\hline
5.2 & Default & Moderate & Uppercut & Yes & Not used\\
\hline
5.2 & Runner-up & Very defensive & Crouch punches & No & Not used \\
\hline
5.2 & Winner &  Defensive &  Crouch kicks & Yes & Used\\
\hline
6.0 & Default & Aggressive  & Heavy punch & No & Not used\\
\hline
6.0  & Runner-up & Aggressive & Punch \& kick & No & Used\\
\hline
6.0  & Winner & Defensive & Uppercut & No & Rarely used\\
\hline
\end{tabular} }
\end{center}
\end{table} 

\subsubsection{DareFightingICE 5.2}
The behavior of the Blind DL AIs trained on DareFightingICE version 5.2 is described in this section.\par

{Default Sound Design --}
For the default sound design, the best encoder is FFT. This Blind DL AI tends to play a moderate role in terms of attack or defense. It tries to jump to avoid the fireball attack from the opponent AI. It attacks by jumping and trying to hit the opponent with kicks. The best move this Blind AI does is that it counters MCTSAI65's jump attacks by performing the uppercut move. This uppercut move seems to be the most used move by this AI.\par

{Runner-up Sound Design --}
 For the runner-up sound design, the best encoder is 1D-CNN. This Blind DL AI plays a very defensive style. It tends to not move from its location much. It lets the opponent come closer and then attacks by crouching and punching. This combo of crouch punches seems to be the most used attack by this AI. It seems that through training, the AI has learned that crouch punching is effective against MCTSAI65. Although it does not try to avoid the fireball attack from the opponent, it can still overwhelm the opponent by trapping them in a corner and spamming crouch punches. \par

{Winner Sound Design --}
For the winner sound design, the best encoder is 1D-CNN. Like the runner-up sound design AI, this AI tends to play defensively. It does not move much apart from some jump attacks. Its favorite combination of attack is crouch kicks as it is the most used attack by the AI. This AI is the strongest among all of the Blind DL AIs trained. It tries to avoid the fireball attack by jumping. Also, it counters MCTSAI65's jumping attack with its own jumping attack, or if it has enough energy, it performs the lighting uppercut. This AI is undefeated in testing.\par

\subsubsection{DareFightingICE 6.0}
The behavior of Blind DL AIs trained on DareFightingICE version 6.0 is described in this section. \par

{Default Sound Design --}
For the default sound design in DareFightingICE version 6.0, the best encoder is FFT. This Blind DL AI is the weakest among the selected AIs. It tends to play a more aggressive role. It attacks by jumping toward the opponent. The most used attack by this AI is the heavy punch. It tends to do a quick sit-stand movement repeatedly. It also does not try to avoid the fireball attack. The uniqueness of this AI is that it performs the throw attack which lifts the opponent in the air. \par

{Runner-up Sound Design --}
For the runner-up sound design in DareFightingICE version 6.0, the best encoder is FFT. This Blind DL AI is the best for DareFightingICE version 6.0. Its behavior is similar to the default sound design AI of DareFightingICE version 6.0. The difference is that this AI tends to attack slowly as compared to the other Blind DL AIs. The most used attack combination of this AI is a punch followed by a kick. It also does the throw move and the lighting uppercut move. It does not try to avoid the fireball attack.\par

{Winner Sound Design --}
For the winner sound design in DareFightingICE version 6.0, the best encoder is 1D-CNN. This Blind DL AI is also defensive. Its most used attack is the uppercut. This AI tries to defend itself from MCTSAI23i's jump attacks with either uppercuts or jumping punches. It scarcely uses the lighting uppercut attack. It also does not try to avoid the fireball attack. \par


\section{Conclusions}
This paper presented a rule-based adaptive background music (BGM) system that consists of five different instruments playing a classical music piece called “Air on G-String.” The proposed adaptive BGM adapts by changing the volume of the instruments. Each instrument is connected to a different element of the game. The paper also showed that the performance of a deep reinforcement learning AI using only audio data as its input, called Blind DL AI, improved while playing with the adaptive BGM as compared to playing without it. We believe that both players with and without vision will also be able to use the same information given by the proposed adaptive BGM to their advantage, just like the Blind DL AI does. \par

In the future, we are planning to use a deep learning approach to the adaptation of BGM. This is because in the current work, we used the performance of the Blind DL AI when fighting against a version of a Monte-Carlo tree search AI to assess our adaptive BGM and we did not do any aesthetic evaluation of it. We believe that the adaptive BGM using deep learning would result in achieving a better performance from the Blind DL AI as compared to the rule-based, as we plan to use the Blind DL AI's performance as a part of the reward function. As for aesthetic evaluation, since our current work is a rule-based adaptive BGM decided the rules taking into account the resulting BGM's aesthetics; However, by using a deep learning approach we anticipate that the aesthetic evaluation of the BGM would be needed. \par

\end{document}